%% file: main.tex
\documentclass{article}

\usepackage{float}
\usepackage{microtype}
\usepackage{graphicx}
\usepackage{tabularx}
\usepackage{subfigure}
\usepackage{booktabs} 
\usepackage{xcolor}

\usepackage{multirow}

\usepackage{hyperref}
\usepackage{graphicx,wrapfig,lipsum}

\usepackage{paralist, tabularx}

\usepackage{lineno}

\usepackage[accepted]{icml2021}

\icmltitlerunning{Neural Multigrid Memory For Computational Fluid Dynamics}

\begin{document}

\twocolumn[
\icmltitle{Neural Multigrid Memory For Computational Fluid Dynamics}



\icmlsetsymbol{equal}{*}

\begin{icmlauthorlist}
\icmlauthor{Duc Minh Nguyen}{equal,Ecole}
\icmlauthor{Minh Chau Vu}{equal,VNU}
\icmlauthor{Tuan Anh Nguyen}{Microsoft}
\icmlauthor{Tri Huynh}{Google}
\icmlauthor{Nguyen Tri Nguyen}{Littelfuse}
\icmlauthor{Truong Son Hy}{UCSD}
\end{icmlauthorlist}

\icmlaffiliation{Ecole}{Ecole Polytechnique, Paris, France}
\icmlaffiliation{VNU}{University of Engineering and Technology, Vietnam National University, Hanoi, Vietnam}
\icmlaffiliation{Microsoft}{Microsoft / OpenAI, California, USA}
\icmlaffiliation{Google}{Google, California, USA}
\icmlaffiliation{Littelfuse}{Littelfuse, Illinois, USA}
\icmlaffiliation{UCSD}{Halıcıoğlu Data Science Institute, University of California, San Diego, USA}

\icmlcorrespondingauthor{Truong Son Hy}{tshy@ucsd.edu}

\icmlkeywords{Computational fluid dynamics, multiscale memory, temporal architecture}

\vskip 0.3in
]



\printAffiliationsAndNotice{\icmlEqualContribution} 

\input{Section/abstract}
\input{Section/introduction}

\begin{figure*}[h!]
  \includegraphics[width=\linewidth]{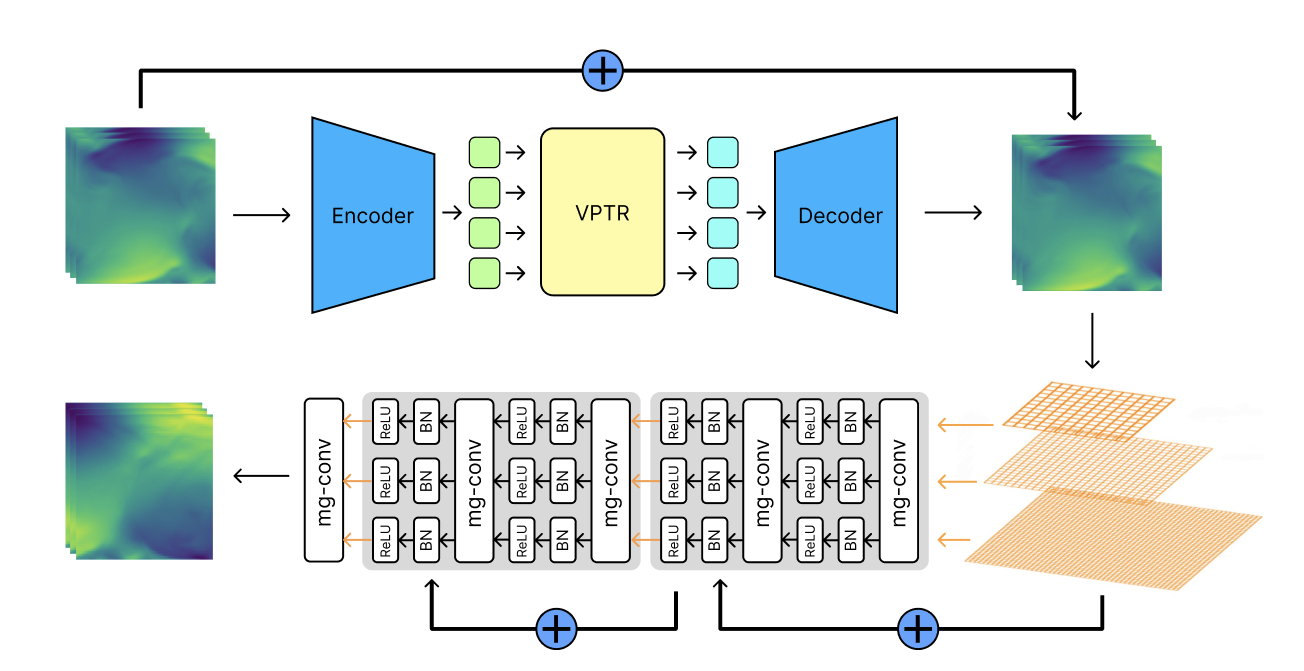}
  \caption{\textbf{Architecture of MGxTransformers.} The model is composed of Video Prediction Transformer (VPTR) and a feed-forward network of Multigrid Convolutional Network (MgNet).}
  \label{fig: Architecture}
\end{figure*}

\input{Section/related_work}

\input{Section/method}

\input{Section/experiments}
\input{Section/conclusion}


\bibliography{main}
\bibliographystyle{icml2021}

\newpage
\appendix
\onecolumn
\input{Section/appendix}


\end{document}

%% file: Section/abstract.tex
\begin{abstract}

Turbulent flow simulation plays a crucial role in various applications, including aircraft and ship design, industrial process optimization, and weather prediction. In this paper, we propose an advanced data-driven method for simulating turbulent flow, representing a significant improvement over existing approaches. Our methodology combines the strengths of Video Prediction Transformer (VPTR) \cite{ye2022vptr} and Multigrid Architecture (MgConv, MgResnet) \cite{Ke_2017_CVPR}. VPTR excels in capturing complex spatiotemporal dependencies and handling large input data, making it a promising choice for turbulent flow prediction. Meanwhile, Multigrid Architecture utilizes multiple grids with different resolutions to capture the multiscale nature of turbulent flows, resulting in more accurate and efficient simulations. Through our experiments, we demonstrate the effectiveness of our proposed approach, named MGxTransformer, in accurately predicting velocity, temperature, and turbulence intensity for incompressible turbulent flows across various geometries and flow conditions. Our results exhibit superior accuracy compared to other baselines, while maintaining computational efficiency. Our implementation in PyTorch is available publicly at \url{https://github.com/Combi2k2/MG-Turbulent-Flow}.
\end{abstract}

%% file: Section/introduction.tex
\section{Introduction} \label{sec:introduction}
Fluid dynamics simulation is crucial in various fields, including aerospace engineering, automotive engineering, chemical engineering, and environmental engineering. However, computational fluid dynamics faces challenges due to high computational costs and the complexity of simulating the Navier-Stokes equations accurately.

Inspired by the success of many fully data-driven deep learning models in computer vision \cite{Yao_Wu_Ke_Tang_Jia_Lu_Gong_Ye_Li_2018}, \cite{NIPS2016_03afdbd6} and the appearance of a new attention and memory mechanism, Multigrid Neural Memory \cite{pmlr-v119-huynh20a}, we have developed another data-driven approach for solving computational fluid dynamics problems. Although the specified memory structure retains the multiscale structure in computational fluid dynamics (CFD), combining both approaches has been shown to be effective.

To verify the strength of our proposal, we first create MgNet \cite{Ke_2017_CVPR}, used by the Multigrid Neural Memory \cite{pmlr-v119-huynh20a}. Additionally, due to the rise of certain transformer approaches in computer vision \cite{yuan2021hrformer}, \cite{ye2022vptr}, we attempt to integrate Transformer layers into the architecture to enable parallel computation. These models are evaluated in predicting flow velocity up to $50$ time steps into the future.

In summary, our contributions include proposing a fully data-driven model for solving computational fluid dynamics problems and improving the efficiency of this approach by incorporating Transformer and Multiscale architectures.

%% file: Section/related_work.tex
\section{Related work}

\paragraph{Spatiotemporal forecasting.}{
Predicting spatiotemporal dynamics is crucial in various fields, such as physics, economics, and neurology. Conventional physics-based differential equations \cite{Izhikevich:2007p75} have been used to model system dynamics but are hard to solve due to sensitivity to initial conditions. Recently, data-driven models using deep learning techniques have been applied to spatiotemporal forecasting, although incorporating physical knowledge into these models can be difficult and modeling turbulence remains a significant challenge for predicting turbulent flow. Despite these challenges, deep learning models have great potential for improving spatiotemporal forecasting.}

\paragraph{Video prediction.}{Our work is also related to future video prediction \cite{Zhai_2022_CVPR}, \cite{NIPS2015_07563a3f}, \cite{pmlr-v119-huynh20a}. Conditioning on the observed frames, video prediction models are trained to predict future frames. There are 2 main problems with previous approaches. The first problem is that many of these models are trained on natural videos with complex noisy data from unknown physical processes which causes difficulty in explicitly learning physical principles in the model. The second problem is the slow inference as the inherited nature of recurrent methods. Hence, some of these techniques are perhaps under-suited for our application.}


\paragraph{Multiscale modeling.}{Various multiscale modeling techniques have been proposed to simulate turbulent flow, including large eddy simulation (LES), hybrid RANS-LES \cite{article}, and particle-based methods. These methods differ in how they address the small-scale features of turbulence. While they have demonstrated positive outcomes, they pose challenges such as computational expenses and modeling the interactions between different scales. Researchers are still working on developing more effective and precise multiscale modeling approaches for simulating turbulent flow.}

\paragraph{Physics-informed Deep Learning.}{
Physics-informed deep learning (PIDL) has recently gained attention for its ability to combine physics-based models and data-driven approaches, by incorporating prior physical knowledge into the loss function of a deep neural network or designing the network to preserve physical properties \cite{Wang2020TF}. PIDL has been applied to various fields, including fluid mechanics, to simulate the dynamics of turbulent flows and estimate physical parameters from observational data. Although PIDL is promising for improving simulations in various fields, it has some limitations, such as the need for careful tuning of hyperparameters and the challenge of including complex physics in the loss function. Nonetheless, PIDL shows potential for significant advancements in the field of fluid mechanics and beyond.}

%% file: Section/method.tex
\section{Method} \label{sec:method}

We propose a data-driven method that replicates Multigrid Neural Memory \citep{pmlr-v119-huynh20a}. In our approach, we notice a chance to parallelize the memory encoding process by replacing ConvLSTM layers by one Transformer layer that recently succeeded in Natural Processing Language. Following that, we choose Video Prediction Transformers, which is recently proved to have more efficient computational cost than traditional Vision Transformer. We observe that the result model can can perform equally to TF-Net \cite{Wang2020TF} which is the current best model that serve this particular purpose.

\subsection{Video Prediction Transformers (VPTR)}

VPTR is an efficient alternative to traditional Vision Transformers, offering comparable performance with improved computational efficiency. It utilizes self-attention mechanisms to capture global dependencies, enabling effective modeling of long-range relationships. However, applying VPTR directly to turbulent flow may compromise interpretability since it operates on a compressed latent space that lacks the ability to effectively capture the inherent multiscale characteristics.

The architecture illustrated in Figure [\ref{fig: Architecture}] showcases the upper section of VPTR, featuring an additional residual connection. The objective is to leverage the learned representations from preceding layers and merge them with the extracted features in subsequent layers. This fusion enhances the autoencoder's capability for feature extraction, resulting in more expressive and adaptable representations.

The detailed implementation of this layer follows the instruction of \cite{ye2022vptr}. The input is the representation of the turbulent flows of the shape $(N, T, C, H, W)$ and the output has the shape $(N, T, d_{model}, H, W)$.

\subsection{Multigrid Convolutional Network (MgNet)}

In our design, we integrate a multigrid variant of CNNs and ResNets, as proposed by \cite{Ke_2017_CVPR}, which inherently exhibit attentional behavior. This architecture consists of multiple convolutional layers operating on different grid sizes. The outputs are concatenated and fed into a final layer for classification or segmentation. MgNet, in our approach, can performs equivalently as an U-Net or Convolutional Feed Forward Net and is designed to capture features at multiple scales.

The layer's input is a pyramid $\mathcal{X} = \{z^x_j\}$, with $j$ representing the pyramid level, obtained by reshaping the output of previous layers (VPTR) into a shape of $(N, T \cdot d_{model}, H, W)$. It is then downsampled and upsampled to create a structured input at multiple scales, as depicted by the orange section in Figure [\ref{fig: Architecture}]. We modify the resolution by a factor of two in each spatial dimension when transitioning between pyramid levels.

The resulting output of this layer remains in the form of a pyramid but now consists of a single level. This level has a specific size of $N\times (T' \cdot C) \times H\times W$, where $T'$ represents the number of frames from the future that we aim to predict. By reshaping this level, we obtain the ultimate prediction of the model.

%% file: Section/experiments.tex
\section{Experiments} \label{sec:experiments}

We investigate the performance of our model by testing its learning ability over 2 datasets that are governs by \href{https://en.wikipedia.org/wiki/Navier%E2%80%93Stokes_equations}{Navier-Stokes Equations}.

We develop a framework to train our model and baselines given the shape of one frame in the sequential data. Here, we employ the PyTorch deep learning library \cite{NEURIPS2019_bdbca288} and set up the same hyper parameters for each model:
\begin{compactitem}
    \item Adam optimizer \cite{kingma2014adam},
    \item Initial learning rate $= 10^{-4}$,
    \item Number of training epochs $= 100$.
\end{compactitem}
Our source code is available at \url{https://github.com/Combi2k2/MG-Turbulent-Flow}.


\begin{table*}[!h]
\begin{center}
\begin{tabular}{ c  c  c  c  c }
\toprule
\multirow{2}{*}{\textbf{Method}} & \multicolumn{2}{c}{\textbf{Rayleigh–Bénard Convection}} & \multicolumn{2}{c}{\textbf{2D Random Flow}}  \\ 
\cmidrule(lr){2-3} \cmidrule(lr){4-5}
& \textbf{MSE $\downarrow$} & \textbf{Divergence $\downarrow$} & \textbf{MSE $\downarrow$} & \textbf{Divergence $\downarrow$} \\
\midrule
ConvFFN & 0.12393 & 0.00115 & 6.163 $\times 10^{-3}$ & 0.207 $\times 10^{-3}$ \\ 
U-Net \citep{10.1007/978-3-319-24574-4_28} & 0.00473 & 0.00762 & 0.764 $\times 10^{-3}$ & 0.271 $\times 10^{-3}$ \\
C-LSTM \citep{NIPS2015_07563a3f} & 0.01279 & 0.00689 & 0.176 $\times 10^{-3}$ & 0.002 $\times 10^{-3}$ \\
TF-Net \citep{Wang2020TF} & 0.05341 & 0.00248 & 0.040 $\times 10^{-3}$ & 0.006 $\times 10^{-3}$ \\
FNO2d \citep{li2021fourier} & 0.00627 & 0.00680 & 0.281 $\times 10^{-3}$ & 0.097 $\times 10^{-3}$ \\ 
\midrule
\textbf{MNM} \citep{pmlr-v119-huynh20a} & 0.01051 & 0.00815 &  12.79 $\times 10^{-3}$ & 6.483 $\times 10^{-3}$ \\
\textbf{MGxTransformer} & 0.00704 & 0.00791 & 0.255 $\times 10^{-3}$ & 0.145 $\times 10^{-3}$ \\
\bottomrule

\end{tabular}
\caption{\label{tab:result} Results on 2 datasets, measured by \textbf{MSE} and \textbf{Divergence} metrics.}
\end{center}
\end{table*}

\subsection{Evaluation Metrics}

To evaluate the effectiveness of our models, we measured the accuracy of their predictions with the following two metric functions:
\begin{compactitem}
\item \textbf{Mean Square Error}: We calculate the MSE of all predicted values from the ground truth for each pixel.
\item \textbf{Divergence Loss}: Since we investigate incompressible turbulent flows in this work, which means the divergence, $\nabla w$, at each pixel should be zero, we use the average of absolute divergence over all pixels at each prediction step as an additional evaluation metric.
\end{compactitem}

\subsection{Rayleigh–Bénard convection (RBC)}
\begin{figure}[h!]
  \includegraphics[width=\linewidth]{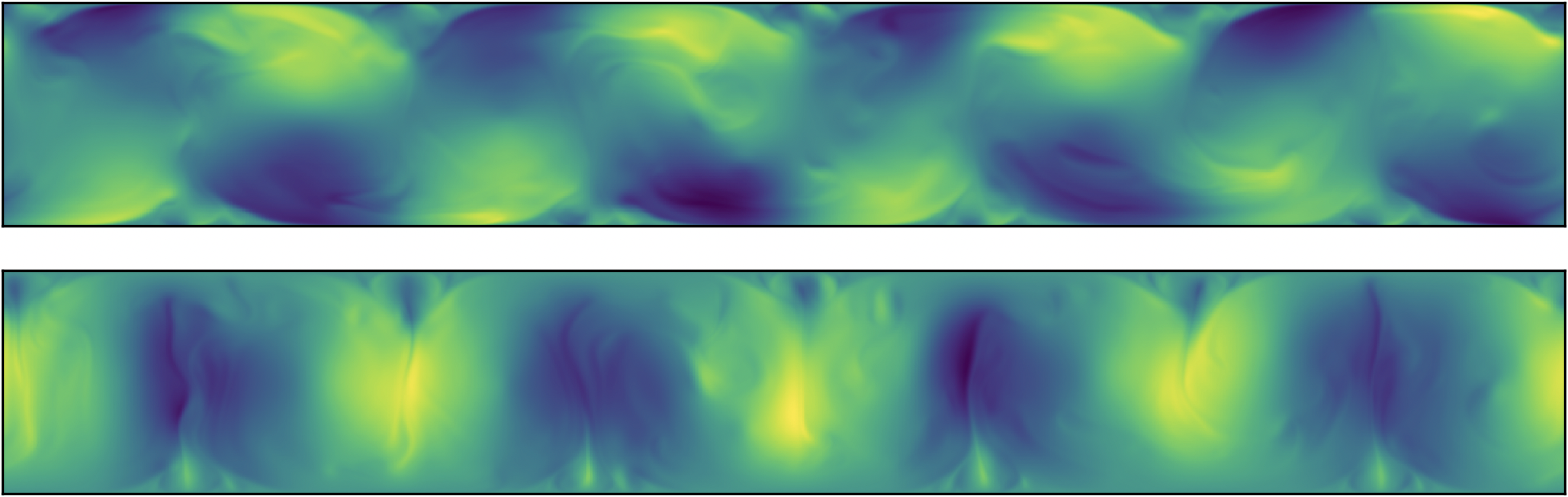}
  \caption{A snapshot of the Rayleigh-Bénard convection flow, the velocity fields along $x$ direction (top) and $y$ direction (bottom). The spatial resolution is 1,792 x 256 pixels.}
  \label{fig: RBC-snapshot}
\end{figure}

This is a model for turbulent convection, with a horizontal layer of fluid heated from below so that the lower surface is at a higher temperature than the upper surface, governed by Navier-Stokes Equations. The dataset comes from two dimensional turbulent flow simulated using the Lattice Boltzmann Method. We use only the velocity vector fields, where the spatial resolution of each frame is $1,792\times 256$ (see Figure [\ref{fig: RBC-snapshot}]). Each image has two channels, one is the turbulent flow velocity along $x$ direction and the other one is the velocity along $y$ direction. The physics parameters relevant to this numerical simulation are: Prandtl number = $0.71$, Rayleigh number = $2.5 × 10^8$ and the maximum Mach number = $0.1$. We use $1,000$ images for our experiments. The task is to predict the spatiotemporal velocity fields up to $50$ steps ahead given $16$ initial frames.

We divided each $1,792\times256$ image into $7$ square sub-regions of size $256\times256$, then downsample them into $64\times64$ pixels sized images. We use a sliding window approach to generate $9,870$ samples of sequences of velocity fields. Here we use only $3,000$ training samples, and $1,000$ validation samples and $1,000$ test samples. The DL model is trained using back-propagation through prediction errors accumulated over multiple steps.

\subsection{2D Random Flow}

The 2D Random Flow dataset utilized in this experiment is derived from the PDEBench benchmark \citep{takamoto2022pdebench}, a comprehensive resource for machine learning experiments in physics \citep{takamoto2022pdebench}. For this particular experiment, we focused on the random flow dataset within the 2D Computational Fluid Dynamics (CFD) sections. The dataset was generated through simulations of the compressible Navier-Stokes equation, considering a shear viscosity of $\eta = 0.01$ and a bulk viscosity of $\zeta = 0.01$. These parameters enable the modeling of fluid behavior in a computationally efficient manner. The dataset comprises a total of 10,000 samples, each of which contains $21$ frames. Each frame consists of two channels representing the $x$-velocity and $y$-velocity components of the fluid flow. The resolution of each frame is set to $128 \times 128$, providing a sufficiently detailed representation of the flow characteristics. To make the dataset ready for testing, we resized the samples to dimensions of $(21, 2, 64, 64)$, where the values indicate the number of timesteps, channels, image height, and image width, respectively. Subsequently, we divided these downscaled samples into three sets: 3,000 for training, 1,000 for validation, and 1,000 for testing. For our model evaluations, we performed tests on 5,000 of these test samples across all models.

\subsection{Result and Observation}

Table \ref{tab:result} presents the evaluation metrics, namely \textit{Mean Squared Error} (MSE) and \textit{Divergence}, for a single prediction step. In terms of the one-step prediction performance, MGxTransformers demonstrates strong potential by outperforming most of the baselines. However, when employing autoregressive methods, our model falls short compared to the TF-Net and FNO2d baselines. Furthermore, regarding divergence loss, our models exhibit a slight performance deficit. This can be attributed to the absence of normal physical laws within the model's structure. The lack of physical reflection within the architecture is the underlying cause of this limitation. Figure \ref{fig:result} visually represents the evolution of the evaluation metrics over a span of $100$ prediction steps, employing an autoregressive method. Notably, our model exhibits a diminishing performance over time, primarily attributed to its inability to preserve essential physical properties and image details within a single prediction step. Nonetheless, these observations underscore the potential efficacy of incorporating physical regularizers as a promising avenue for enhancing our model in future iterations.

%% file: Section/conclusion.tex
\section{Conclusion} \label{sec:conclusion}

We introduce a fully data-driven approach for learning the behavior of turbulent flow. Our methodology incorporates a Transformer layer that effectively captures the spatio-temporal interactions inherent in turbulent flow, akin to a variant of Conv-LSTM. Additionally, the inclusion of MgNet layer facilitates the simulation of the multiscale structure characteristics of fluid dynamics. The end-to-end training pipeline aims to achieve the optimal combination of these components. To comprehensively evaluate the performance of our proposed model, MGxTransformers, we conduct extensive comparisons against various baselines, employing different metrics to assess their respective strengths and limitations. A significant contribution of this research lies in the novel concept of modeling the multiscale structure of turbulent flow through purely deep learning methods. Moving forward, our future work entails the incorporation of physical regularizers such as divergence and turbulence kinetic energy. Moreover, we aim to explore methods that directly integrate the modeling of multiscale structures within the Transformer layers. These endeavors collectively aim to enhance the accuracy and fidelity of deep learning models in the context of turbulent flow analysis.
\newpage

%% file: Section/appendix.tex
\section{Appendix} \label{sec:appendix}

\begin{figure}[h]
    \centering
    \includegraphics[width=0.4\linewidth]{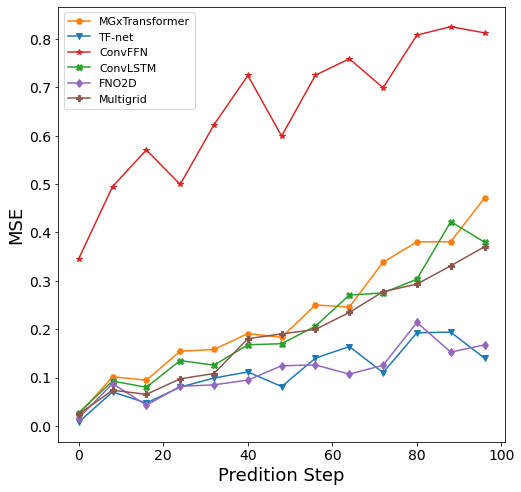}
    \includegraphics[width=0.4\linewidth]{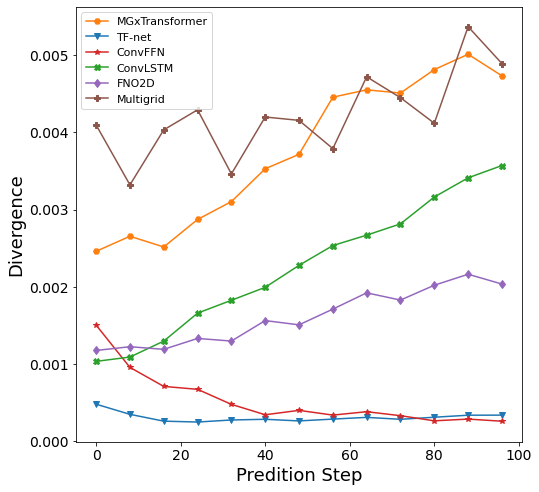}
    \caption{\label{fig:result}\textbf{Mean square errors} (left) and \textbf{mean absolute divergence} (right) of different models’ predictions at varying steps in RBC experiment.}
\end{figure}

Figure \ref{fig:result} shows the mean square errors and mean absolute divergence of different models’ predictions at varying steps in RBC experiment. Multiscale models perform reasonably well in comparison with other competitive baselines accross prediction steps. 

\begin{figure}[h]
    \centering
    \includegraphics[scale=0.42]{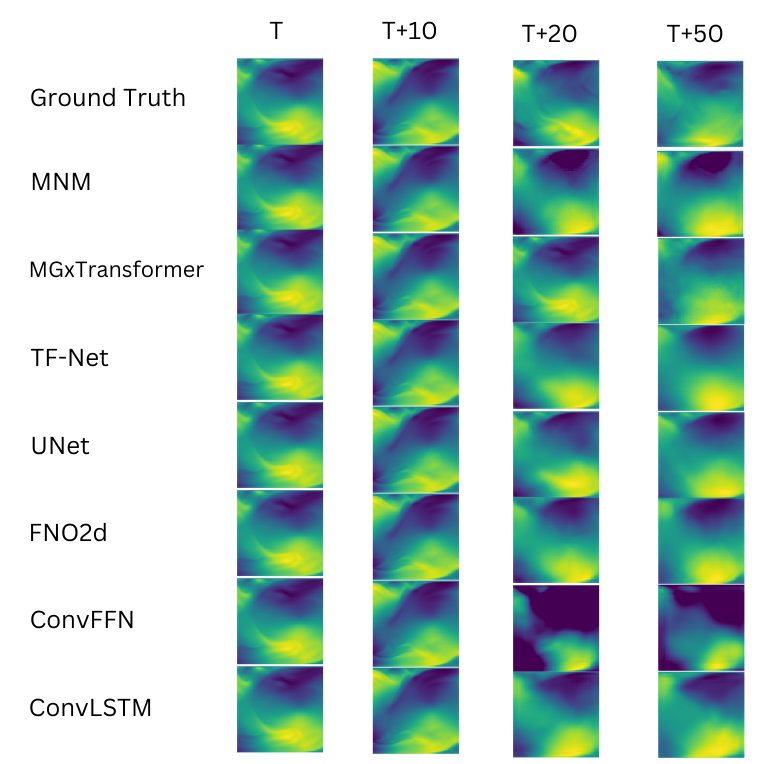}
    \caption{Prediction of all baselines on the Rayleigh–Benard convection dataset upto 50 frames in the future.}
    \label{fig:50_frames}
\end{figure}

We ran all the baselines using 16 input images from the Rayleigh-Bénard convection dataset and employed the sliding window technique to predict up to 50 frames into the future. The results are shown in Figure \ref{fig:50_frames}. In the figure, the names of the baselines are positioned on the left-hand side, while the prediction timesteps are placed at the top. The notation ``T'' in the figure represents the first predicted time frame, indicating that we used 16 frames from T-16 to T-1 as input for the baselines.

